# Magnetic-attractive-interaction-induced superconductivity in metals

Jinhuan Jiang*

**The microscopic BCS theory has described the conventional superconductors successfully. However, it is still not clear when the superconductivity occurs and why the resistance is zero. Also, there is no simple formula to calculate the superconducting temperature Tc, and it is not well understood about how to improve Tc. Therefore, in this paper a microscopic theory of superconductivity caused by magnetic-attractive-interactions in metals is presented, which are based on some idealized assumptions. (1) The magnetic-attractive-interaction between opposite-spin electrons leads to superconductivity. Magnetic-attractive-interactions happen only when the electron velocity reaches up to $10^6$ m / s and the distance between two electrons is about $2 \times 0.0529$ nm. Only electrons near the Fermi surface can participate into the superconductivity, and the number of superconducting electrons is about $10^{-5}$ of the number of free electrons. (2) Resistance is redefined as the ratio of magnetic flux $\Phi_m$ to electric quantity q, i.e., $R=\Phi_m/q$. The resistance equals to zero, R=0, when the magnetic flux $\Phi_m=0$. This is the reason why superconductors show Meissner effect. The external magnetic field would destroy $\Phi_m=0$ and there is a critical magnetic field Bc. (3) Tc is determined by the magnetic-attractive-potential-energy $\varepsilon_p$ and the ratio of the superconducting electrons to total electrons. A simple formula to estimate Tc is derived and it is related not only to the electron density $n$ and Bc, but also to electronic degrees of freedom $i$. For most metals, these theoretical Tc values are in good agreement with the experimental results. (4) According to our theory, there are two ways to increase Tc. The Tc can be increased with increasing the ratio of the superconducting electrons to total electrons, i.e., decreasing the total free electrons or/and increasing the superconducting electrons. If more electrons can be pumped onto the Fermi surface by illuminating, applied a voltage or high pressure, the number of superconducting electrons can be increased, as well as $T_c$. Mankowsky realized the transient room-temperature superconductivity in YBCO by illuminating in 2014. Bilayer magic-angle graphene superconductor was realized by Cao via applied a voltage in 2018. In 2020, room-temperature superconductor was reported under high pressure for the C-S-H system. Another way to improve Tc is to reduce electronic degrees of freedom $i$. Tc in 3-dimensional (3D) Al metal was 1.19 K, and its Tc may be increase to 2.06 K for 1D case. By forming Al "super-atoms" to reduce $i$, Tc was further increased experimentally to more than 100 K in 2015. The maximum Tc of Nb can reach to 377.6 K when 100 Nb atoms form a "super-atom". Based on our theory and the above experiments, we believe that room-temperature superconductors can be realized in Li by applied a voltage at atmospheric pressure since the number of its total electrons is the third lowest among all the elements in the Periodic table.**

In 1911, Onnes found near zero resistance in mercury at 4.2 K[1]. The temperature is called the superconducting transition temperature Tc, where a superconductor loses its resistance. In 1914, Onnes found that the superconducting state can be destroyed by an external magnetic field[2]. When the external magnetic field is greater than the critical magnetic field Bc, electrical resistance suddenly appears, and thus the superconducting state is converted to the normal state. In 1934, Gorter and Casimir proposed a two-fluid model to explain the superconductivity firstly[3]. The key point of this model is that there are two types of electrons in superconductors: normal electrons ($n_n$) and superconducting electrons ($n_s$); the free electron density $n = n_n + n_s$. In 1956, Cooper made an important step in establishing the microscopic theory for superconductivity[4]. He proved that the paired electrons near the Fermi surface (called Cooper pairs) can form bound states, no matter how weak the effective attraction between electrons is. In 1957, Bardeen, Cooper and Schrieffer developed a microscopic theory for superconductivity, which was called BCS theory(5). According to the BCS theory, Cooper pairs are formed by electron-phonon attractive interaction, and they are Boson. The Bose-Einstein condensation of Cooper pairs can lower the energy of the system, making the material superconducting. The decrement of the total energy in the system is called the Bose-Einstein condensation energy Ec. The condensation energy Ec is the sum of the bound energy of all Cooper pairs, i.e., $E_c = n_s \varepsilon/2$, where ε is the bound energy of a Cooper pair. The bound energy is twice that of the BCS superconducting gap Δ, i.e., ε=2Δ and thus we gets $E_c = n_s \Delta$. This energy gap Δ is an important energy scale, which can determine Tc. For an ideally isotropic s-wave superconductor, the energy gap is determined as Δ= $3.52 k_B T_C/2$ experimentally. Cooper pairs are found to exist in all kinds of superconductors, and the breakdown of Cooper pairs is the main reason to destroy superconductivity(6).

* College of Physics and Optoelectronics, Faculty of Science, Beijing University of Technology, Beijing, China, 100124，
E-mail: jiangjh@bjut.edu.cn

The maximum superconducting transition temperature predicted by BCS theory is Tc=40 K, which is called the McMillan limit. It was experimentally found that the critical temperature Tc for $Nb_3Ge$ was 23.2 K(7), which was the highest record until 1985. In 2015, Halder found that a single Al atom showed superconductivity at~1K, but the "super-atom" of Al appeared superconductivity at temperatures up to 100 K, breaking the 40 K limit of BCS theory(8). "Super-atom" is composed of a group of atoms with equal spacing, which is characterized as a single atom. In 2017, the maximum Tc of the new B-doped Q-carbon material was 57 K(9), which also breaks the 40 K limit.

The microscopic BCS theory of superconductivity has described the conventional superconductors successfully. However, it is still not clear when the superconductivity occurs and why the resistance is zero. There is no simple formula to calculate the superconducting temperature Tc, and it is not well understood about how to improve Tc. High Tc in oxide superconductors which is beyond the Mcmillan limit of 40 K cannot be explained. Therefore, in this paper we proposed a theory for the superconductivity induced by magnetic-attractive-interactions in metals, and it is based on the following idealized assumptions:

(1) Magnetic-attractive-interactions between opposite-spin electrons leads to the formation of Cooper pairs and the appearance of superconductivity, and all other kinds of interactions can be omitted. Magnetic-attractive-interactions happen only when the electron velocity reaches up to $10^6$ m / s and the distance between two electrons is about 2 × 0.0529 nm, and these electrons would form Cooper pairs. This magnetic-attractive-interaction between electron spins leads to superconductivity. The corresponding energy is the magnetic-attractive-potential-energy $E_p$, which equals to the Bose-Einstein condensation energy Ec mentioned above, i.e., $E_p = E_c = n_s \Delta$. The magnetic-attractive-potential-energy $\varepsilon_p$ of a Cooper pair is just the bound energy ε of a Cooper pair, i.e., $\varepsilon_p = \varepsilon = 2\Delta$.

The velocity of electrons near the Fermi surface is close to $10^6$ m/s, therefore, only electrons near the Fermi surface can participate into superconductivity. The number of superconducting electrons is very few, i.e., $10^{-5}$ of free electrons. From 2018 to 2021, Cao discovered that the multilayer-graphene is superconducting by adjusting the magic-angle and applying a voltage(10-12). The magic angle was used to adjust the appropriate distance (about 2 × 0.0529 nm) between electrons, and the applied voltage was used to pump electrons onto the Fermi surface, accelerating the velocity of the electrons up to $10^6$ m/s.

(2) The maximum superconducting electron density $n_{S,max}$ approximately equals to $(T_C/T_F)n$, (at T = 0K), i.e., $n_{s,max} \approx (T_C/T_F)n$. Here *n* is the free electron density, $T_F = \hbar^2 (3\pi^2 n)^{2/3}/2m_e k_B$ is the Fermi temperature. For example, for metal Al, $n_{s,max} \approx (T_C/T_F)n = 1.589 \times 10^{24}/m^3$. Since $J_c = n_s e v_F$, the maximum value of critical current density Jc is calculated as 5.16×$10^7$A/$cm^2$. According to Boltzmann distribution law $N(E) = Ne^{-\frac{E}{KT}}$, the electrons near the Fermi surface are very few, $\Delta N_F/N \approx \Delta E_F/E_F = T_C/T_F$. Therefore, the number of superconducting electrons $n_{S,max}$ is far less than that of free electrons *n*, i.e., $n_{S,max} \ll n$, and only about $10^{-5}$ free electrons can participate into superconductivity. Due to $T_C \approx (n_{s,max}/n)T_F$, the Tc can be increased with increasing the ratio of the superconducting electrons to total electrons, i.e., decreasing the total free electrons or/and increasing the superconducting electrons. This is the first way to improve Tc.

Hydrogen might be the best candidate for high-Tc superconductors since it has the minimum number of the total electrons. In 1968, Ashcroft suggested that a high-Tc superconductor can be realized via metallizing Hydrogen[13]. In 2014, Cui's research group theoretically predicted that the Tc of $(H_2S)_2H_2$ was at around 191-204 K at 200 GPa pressures[14]. In 2015, Drozdov reported a superconducting temperature up to 203 K in $H_3S$ at 155 GPa[15]. In 2019, superconductivity in $LaH_{10}$ was found to be higher than 260 K under 170 GPa pressure[16]. In 2020, the room-temperature superconductivity for C-H-S appeared at 288 K under 267 GPa pressure[17]. Lithium might be the best candidate for low-pressure and high-Tc superconductors, since it is metallic itself, and it does not require a further high-pressure metallization process, and the number of its total electrons is the third lowest among all the elements in the Periodic table. In 2021, Cui's group studied the superconductivity of $Li_6C$, and pointed out a direction to screen low-pressure and high-Tc superconductors[18].

On the other hand, the number of superconducting electrons would increase if more low-energy electrons can be pumped onto the Fermi surface, and thus the Tc can be increased. This pumping process can be realized via. light illumination, applied a voltage or high pressure. In 2014, YBCO was found to exhibit superconductivity for a few picoseconds at room-temperature by pulse laser irradiation[19]. The pulsed laser can provide energy to the electrons, and accelerate them up to a velocity of $10^6$ m/s and finally jump them onto the Fermi surface. In this process, the distance between oxygen and copper atoms in Cu-O chain is shortened by 0.022-0.024 nm[19], the corresponding distance between the two electrons may be also shortened to 2×0.0529 nm.

Based on our theory and the above experiments, we speculate that room-temperature superconductors would be realized in metallic lithium by applying a suitable voltage at atmospheric pressure.

(3) Only doubly-occupied states, singly-occupied (spin up or down) states and empty states should be considered in metals. The electrons in the doubly-occupied states form the superconducting states (or Cooper pairs) while those in the singly-occupied states form the normal states.

(4) Resistance is redefined as the ratio of magnetic flux $\Phi_m$ to electric quantity q, i.e., $R = \Phi_m/q$. Following this redefinition, the resistance equals to zero, R=0, when the magnetic field B = 0, and the magnetic flux $\Phi_m$ =0. In this way, it is easy to explain why superconductors have complete diamagnetism or Meissner effect. Also, it can explain the reason why two electrons of a Cooper pair have opposite momentum and spin. When the momentum and spin are opposite, B=0, and $\Phi_m$=0, thus the resistance equals to zero, R=0, and the local superconductivity appears. The spin and momentum of electrons are not in opposite direction any longer upon heating, leading to destroy the magnetic-attractive-interaction of Cooper pairs(i.e. $\Phi_m \neq 0$), and to transform the superconducting states into normal states. Therefore, there is a critical temperature Tc. The energy that Cooper pairs get from hot source to destroy all Cooper pairs equals to the magnetic-attractive-potential-energy $E_p$, i.e., $E_p = n_S \varepsilon_p/2 = n_S \Delta$. Obviously, Tc is determined by the magnetic-attractive-potential-energy $\varepsilon_p$ of a Cooper pair. And an external magnetic field can also destroy $\Phi_m$=0, the opposite momentum and spin, the magnetic-attractive-interaction of Cooper pairs, and the superconductivity, therefore, there is a critical magnetic field Bc. The energy $E_p$ needed to destroy all Cooper pairs can be figured out by the external magnetic field Bc, i.e., $E_p = n_S \varepsilon_p/2 = B_C^2/2\mu_0 = n_S \Delta$. The superconducting critical temperature Tc also depends on the critical magnetic field $B_c$.

(5) The average kinetic energy of electrons complies with the equipartition theorem of energy. According

to this theorem, the maximum average kinetic energy of the superconducting electrons is $\varepsilon_k = ik_BT_C/2$ ($i = 1,2,3,\cdots$), where $k_B$ is the Boltzmann constant, and $i$ is the electronic degrees of freedom. $i = 1$, for 1D system, and for 2D system, $i = 2$. An increment in temperature can break up the Cooper pairs, and thus destroy the superconductivity. Therefore, the magnetic- attractive-potential-energy $\varepsilon_p$ of a Cooper pair equals to its average kinetic energy, i.e. $\varepsilon_p = 2\varepsilon_k = 2(ik_BT_C/2) = 2\Delta$, and thus the superconducting energy gap is $\Delta= \varepsilon_k = ik_BT_C/2$. For the ideally isotopic s-wave superconductors in metals, $i = 3$, the theoretical value of the superconducting energy gap $\Delta= 3k_BT_C/2$, is close to the experimental results $\Delta= 3.52k_BT_C/2$.

In the next step, we will derive a simple expression of the superconducting transition temperature Tc.

As stated above, only a small number of electrons contribute to the superconductivity in metals, and the number of superconducting electrons reaches to its maximum at T = 0K. The maximum density of the superconducting electrons is

$$n_{S,max} = (T_C/T_F)n \neq n. \tag{1}$$

These superconducting electrons can generate Cooper pairs via magnetic-attractive-interactions, and form bound states to reduce the energy of the system, and turn metals into superconductors at sufficiently low temperatures. The magnetic-attractive-interactions in Cooper pairs will be destroyed by further increasing temperature, leading to a decrement of the number of superconducting electrons. At $T_C$, all Cooper pairs are broken up, and the number of superconducting electrons drops to zero. The energy density E in a superconducting system is the sum of the electron average kinetic energy $\varepsilon_k$ and the magnetic-attractive-potential-energy $\varepsilon_p$

$$E = (-n_S(T)\varepsilon_p/2) + n_S(T)\varepsilon_k = -n_S(T)\Delta + n_S(T)\frac{i}{2}k_BT. \tag{2}$$

For $E \leq 0$, metals will be in the superconducting states, and thus $T \leq (2\Delta/ik_B) = T_C$. This is exactly the critical temperature $T_C$, and $\varepsilon_p = ik_BT_C$. Obviously, Tc is determined by the magnetic-attractive-potential-energy $\varepsilon_p$, however, we know little about $\varepsilon_p$.

An external magnetic field can also destroy Cooper pairs, and decrease the number of superconducting electrons. When a critical magnetic field $B_C$ is reached, all the Cooper pairs are broken up, and the number of superconducting electrons becomes zero. The energy density in a superconducting system under an external magnetic field is

$$E = -n_S(T)\Delta + \frac{B^2(T)}{2\mu_0} + n_S(T)\frac{i}{2}k_BT. \tag{3}$$

For $E \leq 0$, metals will be in the superconducting states, and we get $B \leq \sqrt{2\mu_0 n_S(0)\Delta}= B_C$ at T = 0 K. This is exactly the critical magnetic field $B_C$. Thus,

$$n_{S,max}\frac{i}{2}k_BT_C = \frac{B_C^2}{2\mu_0}. \tag{4}$$

Here one can see that the maximum magnetic-attractive-potential-energy in metals equals the maximum energy of the external magnetic field. Inserting Eq. (1) into Eq. (4), the superconducting critical temperature can be figured out by $i$

$$T_C = \frac{\hbar(3\pi^2)^{1/3}}{k_B\sqrt{2m_e\mu_0}} \times \frac{B_C}{n^{1/6}\sqrt{i}} = 1.56 \times 10^7 \times \frac{B_C}{n^{1/6}\sqrt{i}}. \tag{5}$$

where $i$ is the electronic degrees of freedom, $k_B$ is the Boltzmann constant, $m_e$ is the electron mass, $\hbar$ is the reduced Planck constant, $\mu_0$ is the vacuum permeability. Obviously, $T_C$ is not only related to the critical magnetic field $B_C$ and the electron density $n$, but also related to the electronic degrees of freedom $i$.

The $T_C$ values calculated from Eq.(5) for 21-kinds of metals are in good agreement with the experimental results as shown in the table-1. The degree of freedom $i$ may be determined by the shape and topological properties of the Fermi surface.

| No. | Degrees of freedom $i$ | Metals | Electron density $n$ ($10^{29}/m^3$) | Critical magnetic field $B_C$ ($10^{-4}$T) | Theoretical $T_C$(K) | Experimental $T_C$(K) |
|---|---|---|---|---|---|---|
| 1 | $i=3$ | Al | 1.812 | 99 | 1.19 (If $i=1$, Tc=2.06K; If $i=2$, Tc=1.46K.) | 1.19 |
| 2 | | Sn | 1.482 | 303 | 3.76 | 3.72 |
| 3 | | In | 1.145 | 279 | 3.62 | 3.41 |
| 4 | | W | 3.803 | 1.15 | 0.012 | 0.012 |
| 5 | | Mo | 3.840 | 93 | 0.98 | 0.92 |
| 6 | | Tl | 1.048 | 180 | 2.37 | 2.38 |
| 7 | | Os | 5.692 | 65 | 0.64 | 0.65 |
| 8 | | Hg($\alpha$) | 4.878 | 409 | 4.16 | 4.15 |
| 9 | | Ir | 6.320 | 15 | 0.146 | 0.14 |
| 10 | $i=5$ | Re | 4.646 | 205 | 1.63 | 1.70 |
| 11 | | Ru | 5.859 | 66 | 0.505 | 0.49 |
| 12 | | Ti($\alpha$) | 2.284 | 56 | 0.50 | 0.49 |
| 13 | $i=6$ | Pb | 1.322 | 804 | 7.19 | 7.20 |
| 14 | $i=7$ | Th($\alpha$) | 1.214 | 163 | 1.37 | 1.37 |
| 15 | $i=2$ | Zn | 1.315 | 54 | 0.84 | 0.84 |
| 16 | | Zr | 1.721 | 47 | 0.70 | 0.73 |
| 17 | | Cd | 0.9624 | 30 | 0.49 | 0.52 |
| 18 | $i=1$ | Ga | 1.529 | 55 | 1.18 | 1.1 |
| 19 | $i=12$ | Ta | 2.777 | 823 | 4.6 (If $i=1$, Tc=15.93K.) | 4.48 |
| 20 | | V | 3.522 | 1020 | 5.48 (If $i=1$, Tc=18.96K.) | 5.45 |
| 21 | $i=16$ | Nb | 2.776 | 1950 | 9.44 (If $i=1$, Tc=37.76K; If $i=1/4$, Tc=75.52K; If $i=1/100$, Tc=377.6K.) | 9.26 |

Table-1   Comparison between theoretical and experimental $T_C$ for 21-kinds for metals(20)

Based on a magnetic-attractive-interaction mechanism as derived Eq. (5), $T_C$ becomes higher for a lager critical magnetic field $B_C$, or a smaller degree of freedom $i$, or a smaller total free electrons $n$. Therefore, the second method to improve Tc is to reduce the electronic degrees of freedom $i$. Generally $T_C$ increases with decreasing the dimension of metals. $T_C$ in 1D will be higher than that in 3-dimensional (3D) metal. For example, $T_C$ in 3D Al metal was 1.19 K, and $T_C$ in 2D Al is calculated as 1.46 K, and its Tc may be increase to 2.06 K in 1D. Tc was improved experimentally by forming Al "super-atoms" to reduce $i$ in 2015, and Tc for Al reached to more than 100 K[8]. The Tc of Nb is estimated to be 37.76 K at i =1. The maximum Tc for Nb can reach to 377.6 K when 100 Nb atoms form a "super-atom", i.e. $i = 1/100$.

In conclusion, we proposed a microscopic theory of magnetic-attractive-interaction-induced superconductivity in metals, and it is based on the following idealized assumptions: (1) Magnetic- attractive-interactions between opposite-spin electrons lead to superconductivity; (2) Only a small number of electrons are involved in superconductivity; (3) There are different electronic states, i.e., doubly-occupied, singly-occupied (spin up or down) and empty states; (4) Resistance is redefined as the ratio of magnetic flux $\Phi_m$ to electric quantity q, i.e., R=$\Phi_m$/q. (5) The average kinetic energy of electrons complies with the equipartition theorem of energy. There will be magnetic-attractive- interactions only when electron velocities reach $10^6$ m/s, and the electron distance closes to 2×0.0529nm, therefore, only about $10^{-5}$ free electrons participate in the superconductivity. The resistance equals to zero R=0 when the magnetic flux $\Phi_m$=0. This is the reason why superconductors show the complete diamagnetism. The superconducting critical temperature Tc can be determined by the magnetic-interaction-potential-energy $\varepsilon_p$, and thus a simple formula to estimate $T_C$ was derived. The $T_C$ values calculated from this formula are in good agreement with the experimental results for most metals.

Basically, there are two methods to increase Tc. (1) To pump more electrons onto the Fermi surface. Mankowsky realized the transient room temperature superconductor by pumping more electrons onto the Fermi surface by illuminating in 2014. Graphene superconductor was realized by Cao via applying a voltage to pump electrons onto the Fermi surface in 2018. In 2020, room temperature superconductor was reported in C-H-S under high pressure, where more electrons can be pumped onto the Fermi surface. (2) To reduce the electronic degrees of freedom *i*. For example, Tc in 3D Al was 1.19 K, and its Tc may be increase to 2.06 K for 1D case. By forming Al super-atoms to reduce i, Tc was improved experimentally in 2015, and Tc for Al reached to more than 100 K. The maximum Tc of Nb can reach 377.6 K when 100 Nb atoms form a "super-atom". Based on our theory and the above experiments, we believe that room-temperature superconductors can be realized in Li by applying a voltage at atmospheric pressure.

The new definition of resistance R = $\Phi_m/q$ may be explained the Quantum Hall effect, including the integer quantum Hall effect and fraction quantum Hall effect. Because the magnetic flux and electric quantity are quantized, i.e. $\Phi_m = n(h/e)$ and q = me, the resistance is quantized, R = $\Phi_m/q$ = (n/m)($h/e^2$) = (ν)($h/e^2$), where ν = n/m is the filling factor, and n and m are integers. The magnetic-attractive-interaction-induced mechanism of high-Tc copper oxide superconductors is under investigation.